\documentclass[reprint,twocolumn,aps,prd,superscriptaddress]{revtex4-1}

\usepackage{graphicx}         
\usepackage{amssymb}
\usepackage{amsmath}
\usepackage{amsthm}
\usepackage[colorlinks]{hyperref}
\usepackage{color,rotating}
\usepackage{bbm}
\usepackage{array}
\usepackage{tikz}
\usetikzlibrary{fit, matrix}
\usepackage{colonequals}
\usepackage{dashrule}
\usepackage{ulem}

\definecolor{darkgreen}{rgb}{0,0.6,0}
\definecolor{gray}{rgb}{.7,.7,.7}

\newlength\replength
\newcommand\repfrac{.33}

\newcommand\rulewidth{.6pt}
\newcommand\tdashfill[1][\repfrac]{\cleaders\hbox to \replength{%
  \smash{\rule[\arraystretch\ht\strutbox]{\repfrac\replength}{\rulewidth}}}\hfill}

\newcommand\tdotfill[1][\repfrac]{\cleaders\hbox to \replength{%
  \smash{\raisebox{\arraystretch\dimexpr\ht\strutbox-.1ex\relax}{.}}}\hfill}

\DeclareMathAlphabet{\EuRoman}{U}{eur}{m}{n}
\SetMathAlphabet{\EuRoman}{bold}{U}{eur}{b}{n}

 \graphicspath{{./figures/}}

\def\di{\displaystyle}

\def\bg{\begin{eqnarray}\begin{array}{rcl}\displaystyle}
\def\eg{\end{array} &\di    &\di   \end{eqnarray}}
\def\bm#1{\begin{eqnarray}\begin{array}{#1}\di}
\def\bmo#1{\begin{eqnarray*}\begin{array}{#1}\di}
\def\bml#1#2{\begin{eqnarray}\begin{array}{#1}\label{#2}\di}
\def\bgo{\begin{eqnarray*}\begin{array}{rcl}\displaystyle}
\def\ego{\end{array} &\di    &\di \nonumber  \end{eqnarray*}}

\def\btensor#1#2{\renew\left#1\begin{array}{#2}\di}
\def\brtensor#1#2#3{\ren#3\left#1\begin{array}{#2}}
\def\botensor#1#2{\renew\left#1\begin{array}{#2}}
\def\etensor#1{\end{array}\right#1}

\def\eq#1{(\ref{#1})}

\def\s0#1#2{\mbox{\small{$ \frac{#1}{#2} $}}}
\def\0#1#2{\frac{#1}{#2}}

\def\CN{{\mathcal N}}

\def\s{\sigma}

\def\ren#1{\renewcommand{\arraystretch}{#1}}

\def\renew{\renewcommand{\arraystretch}{1}}

\def\0#1#2{\frac{#1}{#2}}
\def\eq#1{(\ref{#1})}

\newcommand{\UV}{{\small UV}}
\newcommand{\IR}{{\small IR}}

\allowdisplaybreaks

\begin{document}

\title{Local Quantum Gravity}

%
\author{N. Christiansen}

\affiliation{Institut f\"ur Theoretische Physik, Universit\"at Heidelberg,
Philosophenweg 16, 69120 Heidelberg, Germany}
\author{B. Knorr}

\affiliation{Theoretisch-Physikalisches Institut, Universit\"at Jena,
Max-Wien-Platz 1, 07743 Jena, Germany}
\author{J. Meibohm}

\affiliation{Institut f\"ur Theoretische Physik, Universit\"at Heidelberg,
Philosophenweg 16, 69120 Heidelberg, Germany}
\author{J. M. Pawlowski}
\affiliation{Institut f\"ur Theoretische Physik, Universit\"at Heidelberg,
Philosophenweg 16, 69120 Heidelberg, Germany}
\affiliation{ExtreMe Matter Institute EMMI, GSI Helmholtzzentrum f\"ur
Schwerionenforschung mbH, Planckstr.\ 1, 64291 Darmstadt, Germany}
\author{M. Reichert}

\affiliation{Institut f\"ur Theoretische Physik, Universit\"at Heidelberg,
Philosophenweg 16, 69120 Heidelberg, Germany}


\begin{abstract}

  We investigate the ultraviolet behaviour of quantum gravity within a
  functional renormalisation group approach. The present setup
  includes the full ghost and graviton propagators and, for the first
  time, the dynamical graviton three-point function. The latter gives
  access to the coupling of dynamical gravitons and makes the system
  minimally self-consistent. The resulting phase diagram confirms the
  asymptotic safety scenario in quantum gravity with a non-trivial
  \UV{} fixed point.

  A well-defined Wilsonian block spinning requires locality of the
  flow in momentum space. This property is discussed in the context of
  functional renormalisation group flows. We show that momentum
  locality of graviton correlation functions is non-trivially linked
  to diffeomorphism invariance, and is realised in the present setup.

\end{abstract}

\maketitle

{\it Introduction - }
One of the major challenges in theoretical physics is the unification
of the standard model of particle physics with gravity. A very
promising route is the asymptotic safety scenario proposed by Weinberg
in 1976 \cite{Weinberg:1980gg}.  In this scenario, quantum gravity
exhibits a nontrivial ultraviolet (\UV{}) fixed point with a finite
dimensional critical hypersurface, which renders the theory finite
and predictive even beyond the Planck scale.

A large amount of work in the recent years supports the existence of
such a gravitational \UV{} fixed point. A non-perturbative method of
choice for studying the asymptotic safety scenario is the functional
renormalisation group. The simplest approximation in this approach is
given by an RG-improved Einstein-Hilbert truncation. In the latter,
one retains only two couplings, namely Newton's coupling $G_N$ and the
cosmological constant $\Lambda$ \cite{Reuter:1996cp}. Already this
basic truncation exhibits a \UV{} fixed point, and truncations beyond
Einstein-Hilbert, e.g.\
\cite{Christiansen:2012rx,Christiansen:2014raa,Donkin:2012ud,
  Falls:2014tra,Codello:2007bd,Machado:2007ea,Benedetti:2009rx},
provide further evidence for its existence. The implementation of
diffeomorphism invariance and the related interplay of background and
fluctuation fields is addressed in e.g.\
\cite{Manrique:2010am,Christiansen:2012rx,Donkin:2012ud,Christiansen:2014raa,
  Becker:2014qya}. The analysis of the infrared (\IR{}) limit is
subject to the studies
\cite{Donkin:2012ud,Nagy:2012rn,Christiansen:2012rx,Christiansen:2014raa}. The
first smooth, classical \IR{} limit was found in
\cite{Christiansen:2014raa}, delivering a global picture of quantum
gravity.

In this letter we present the first calculation of a genuine dynamical
gravitational coupling. Here we build on the general setup for flows
of fully momentum dependent vertex functions in quantum gravity
developed in \cite{Christiansen:2012rx,Christiansen:2014raa}. This
expansion naturally resolves the physically important difference
between the graviton wave-function renormalisation and the
gravitational couplings. The existence of the \UV{} and \IR{} fixed
points is confirmed in this enhanced approximation, thus providing
further evidence for the asymptotic safety scenario.  Interestingly,
the ultraviolet fixed point exhibits one irrelevant direction, 
along with two relevant ones, in accordance with the hypothesis of a
finite dimensional critical hypersurface. For related results in
$f(R)$ gravity see e.g.\ \cite{Benedetti:2013jk,Falls:2014tra}.

A well-defined Wilsonian block spinning requires locality of the flow
in momentum space. Here we show that the flows of the graviton two-
and three-point functions are local in momentum space. This
non-trivial property is linked to diffeomorphism invariance.  \\[-1ex]

{\it The Functional Renormalisation Group - }
The present computation of correlation functions in quantum gravity is
based on the functional renormalisation group approach to gravity \cite{Reuter:1996cp}. In this
framework momenta smaller than the infrared cutoff scale $k$ are
suppressed by including a momentum dependent mass function
$R_k$. This leads to a scale-dependent effective
action $\Gamma_k[\overline g,\phi]$. The full metric $g= \bar g +h$ is
expanded around a fixed background metric $\overline g$, and the
fluctuation super-field $\phi=(h,c,\overline c)$ comprises the
graviton fluctuations $h$ and the (anti-) ghost fields $c, \overline
c$. The scale-dependence of the effective action is governed by the
flow of $\Gamma_k$, to wit
\begin{align}
  \dot \Gamma_k &= \frac{1}{2} \text{Tr}\left[ \frac{
      1}{\Gamma_k^{(2)}+R_{k}}\dot R_{k} \right]_{hh}\hspace*{-0.3cm}
  - \text{Tr}\left[ \frac{1}{\Gamma_k^{(2)}+R_{k}}\dot R_{k}
  \right]_{\bar c c} \hspace*{-0.15cm} .
\label{eq:gen_flow_eq}
\end{align}
In \eq{eq:gen_flow_eq} we have introduced the notation $\dot
f=\partial_t f$ for $t = \log k/k_{\text{\tiny{in}}}$-derivatives with
some reference scale $k_{\text{\tiny{in}}}$, usually taken to be the
initial scale. The trace implies integrals over continuous and sums
over discrete indices, and \begin{small}$\Gamma^{(2)}_k$\end{small} is the second derivative
of the effective action with respect to the fluctuation fields.

An important issue in quantum gravity is the background independence
of physical observables. We emphasise that in the present framework,
based on an effective action $\Gamma[\overline g,\phi]$, we have the
paradoxical situation that the background independence of observables
necessitates the background dependence of the vertex functions
$\Gamma^{(n)}$ of the dynamical fields. Importantly, an ansatz
$\Gamma[\overline g,\phi]=\Gamma[\bar g+\phi]$ violates background
independence and dynamical diffeomorphism invariance, see e.g.\ 
\cite{Pawlowski:2003sk,Pawlowski:2005xe,Donkin:2012ud}. 

A setup that disentangles the dependence on $\bar g$ and
$\phi$ in terms of a vertex expansion has been constructed in
\cite{Christiansen:2012rx,Christiansen:2014raa}.  In this vein, a
vertex expansion about a flat Euclidean background is used in this
work to self-consistently determine the flow of the
couplings from the two- and three-point functions. \\[-1ex]

{\it Locality - }
The functional renormalisation group is based on the idea of a
successive integration of momentum shells, or, more generally,
spectral shells of spectral values of the given kinetic
operator. Hence, it relies on the distinction of small and large
momentum or spectral modes. A functional RG step implements the
physics of momentum/spectral modes at a given scale $k$ and is
inherently related to local interactions.

Locality in momentum space implies in particular that the flows of
vertices at a given momentum scale $k$ decay relative to the vertex
itself if all momentum transfers (momentum channels) $t_i$ are taken
to infinity. For example, for the four-point vertex we have
$t_1,t_2,t_3$ being the well-known $s,t,u$-channels, with
e.g. $s=(p_1+p_2)^2$. Hence, locality reads schematically
\begin{align}\label{eq:Gninfty} 
\lim_{t_i/k^2 \to \infty }  \0{|\dot 
      \Gamma_k^{(n)}(\mathbf{p})|}{|\Gamma_k^{(n)}(\mathbf{p})|}=0\,, 
\quad {\rm with} \quad \mathbf{p}=(p_1,...,p_n)\,,
\end{align} 
where a projection on one of the tensor structure of the vertex is
implied. For the limit \eq{eq:Gninfty} each diagram in the flow of
a given vertex has an infinite momentum transfer. Thus, the diagrams
are only sensitive to fluctuations far above the cutoff scale.

It is easily proven that \eq{eq:Gninfty} applies to standard
renormalisable quantum field theories in four dimensions including
non-Abelian gauge theories that involve momentum-dependent couplings.
In these theories, the locality property follows from power-counting
arguments.  However, for perturbatively non-renormalisable theories in four dimensions
power counting suggests non-local flows and \eq{eq:Gninfty} must be a
consequence of non-trivial cancellations. In gravity this has been
shown for the graviton propagator
\cite{Christiansen:2012rx,Christiansen:2014raa}. It is also reflected
in the symmetry relation between graviton diagrams contributing to the
Yang-Mills propagator \cite{Folkerts:2011jz}. Moreover, it is easily
verified that a $\phi^4$-theory with a momentum-dependent coupling
such as $\int_x \phi^2\partial^2\phi^2$ does not satisfy the locality
condition \eq{eq:Gninfty}, as no cancellation between tensor
structures is possible. This entails that momentum locality in quantum
gravity is linked to diffeomorphism invariance, and we conjecture that
it is indeed rooted in the latter.

Note that \eq{eq:Gninfty} does not hold, even for quantum field theories which are
perturbatively renormalisable in four dimensions, if some of the channels
$t_i/k^2$ stay finite: the flow always involves diagrams with a finite
momentum transfer. However, those diagrams correspond to IR processes
such as Bremsstrahlung, which is why they do not reflect the UV
behaviour of the theory. In summary the above discussion suggests that
the relation \eq{eq:Gninfty} is a necessary requirement for local
quantum field theories.

\begin{figure}[t]
\centering
\includegraphics[width=8.5cm]{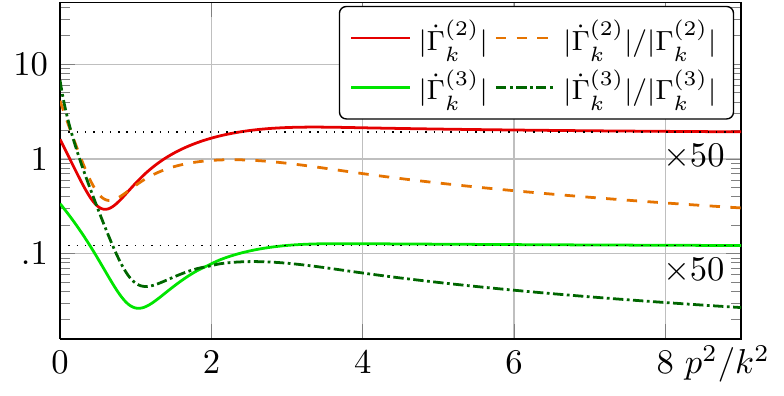}
\caption{Logarithmic plot of the flows \begin{small}$|\dot{\Gamma}^{(2)}_k|$\end{small} and
  \begin{small}$|\dot{\Gamma}^{(3)}_k|$\end{small} (solid red and light green curves) and the
  corresponding ratios \begin{small}$|\dot{\Gamma}^{(2)}_k|/|\Gamma^{(2)}_k|$\end{small} and
  \begin{small}$|\dot{\Gamma}^{(3)}_k|/|\Gamma^{(3)}_k|$\end{small}
  (dashed orange and dash-dotted dark green curves) as functions of
  $p^2/k^2$. The norm refers to the tensor projection discussed below
  \eqref{e:tendec}.  All quantities are evaluated at
  $(g,\mu_h,\lambda_3)=(1,0.1,-0.7)$. The flows are multiplied with 50
  for convenience. The ratios decay with $1/p^2$ for large $p$ since
  the associated flows quickly approach constant values, satisfying
  \eqref{eq:Gninfty}.}
\label{fig:locality}
\end{figure}

In this work, we show (see page 3 \& 4) that \eq{eq:Gninfty} also
applies to the graviton three-point function. Together with the
momentum locality of the two-point function shown in
\cite{Christiansen:2012rx,Christiansen:2014raa} this provides strong
indications for the momentum locality of
RG-gravity. \autoref{fig:locality} depicts the momentum dependence of
the flows for the graviton two- and three-point
functions, \begin{small}$|\dot \Gamma^{(2)}_k|$\end{small}
and \begin{small}$|\dot \Gamma^{(3)}_k|$\end{small}, respectively as
well as the corresponding ratios according to
\eqref{eq:Gninfty}. Since \begin{small}$|\dot
  \Gamma^{(2)}_k|$\end{small} and \begin{small}$|\dot
  \Gamma^{(3)}_k|$\end{small} quickly approach
constants, the ratios decay with $1/p^2$ for large momenta. \\[-1ex]

\begin{figure}[t]
\centering
\includegraphics[width=8.5cm]{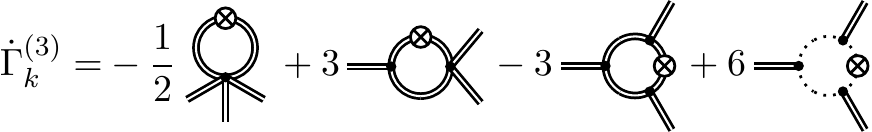}
\caption{Diagrammatic representation of the flow of the three-graviton
  vertex. Double and dashed lines represent graviton and ghost
  propagators respectively, filled circles denote dressed
  vertices. Crossed circles are regulator insertions. All diagrams are
  symmetrised with respect to the interchange of external momenta
  $\mathbf{p}$.}
\label{fig:flow_three_point}
\end{figure}

{\it Flows of Correlation Functions - }
The flow of the three-point function is obtained by three field
derivatives of the flow equation for the effective action
\eqref{eq:gen_flow_eq}. It is depicted in
\autoref{fig:flow_three_point}.  We build on the parametrisation for
vertex functions introduced in
\cite{Fischer:2009tn,Christiansen:2014raa}. Our ansatz is given by
\begin{align}\label{vertex}
  \Gamma_k^{(\phi_1 \dots \phi_n)}(\mathbf{p}) = \left(\prod_{i=1}^n
    \sqrt{Z_{\phi_i}(p_i^2)}\right) G_n^{\frac{n}{2}-1} \,
  \mathcal{T}^{(n)}(\mathbf{p};\Lambda_{n}) \,. 
\end{align}
where $ \mathcal{T}^{(n)}$ is the classical tensor structure of the
vertex to be specified in \eq{tensor_struc}. Other tensor structures
are neglected. Moreover, $Z_{\phi_i}(p_i^2)$ denotes the wave function
renormalisation corresponding to the field $\phi_i(p_i)$ and captures
the renormalisation group running of the vertex. The coupling
$G_n({\mathbf{p}})$ represents the gravitational vertex coupling of
$n$-th order, and $\Lambda_n$ parametrises the momentum independent
part of \begin{small}$\Gamma^{(n)}_k$\end{small}. Note that $G_3$ is the first dynamical
gravitational coupling and its flow is extracted from the graviton
three-point function. In this work we approximate all $G_n$ as one,
momentum-independent coupling, $G_n(\mathbf{p})=G_3$. We shall show that the
momentum-dependence of the vertex is already captured well with that
of the factors $Z^{1/2}_{\phi_i}(p_i^2)$. We also emphasise
that $G_n$, $\Lambda_n$ and $Z_{\phi_i}$ are scale dependent, 
although the subscript $k$ is suppressed here and in the following. 

Finally, the tensor structures $\mathcal{T}^{(n)}$ are obtained from
the classical gauge fixed Einstein-Hilbert action,  
\begin{align}\label{tensor_struc}
  \mathcal{T}^{(n)}(\mathbf{p};
  \Lambda_n)= G_N\,S^{(n)}(\mathbf{p};\Lambda \rightarrow
  \Lambda_n)\,,
\end{align}
and 
\begin{align}\label{EH_action}
S = \frac1{16 \pi G_N} \int \text{d}^4x
\sqrt{g}\;(2\Lambda-R)+S_\text{gf}+S_\text{gh} \,.
\end{align}
In \eqref{EH_action}, $R$ is the Ricci scalar and $\Lambda$ the
cosmological constant. The terms $S_\text{gh}$ and $S_\text{gf}$ are
the usual Faddeev-Popov ghost action and the gauge fixing term, respectively. We
employ a De-Donder-type linear gauge condition in the Landau
limit of vanishing gauge parameter.

The right hand side of the vertex flows usually includes all types of
tensor structures admitted by symmetry irrespective of the ansatz for
the vertices. For the flow equations for the couplings
$\Lambda_n$ and $G_n$ we have to project the tensorial vertex flow
appropriately: we focus on the transverse-traceless parts of the flow,
and reduce all external graviton legs to its spin-2 parts by using
transverse-traceless projectors $\Pi_{\text{TT}}$.  The flow of $\Lambda_n$ 
in \eqref{tensor_struc} is extracted from the momentum independent part of 
the $n$-th vertex flow at $\mathbf{p}=0$. Consequently, we decompose
$\mathcal{T}^{(n)}(\mathbf{p},\Lambda_n)$ into its momentum
independent part $\mathcal{T}^{(n)} (0,1)$ and the part
$\mathcal{T}^{(n)}(\mathbf{p},0)$, which is quadratic in $\mathbf{p}$,
according to
\begin{align}\label{e:tendec}
  \mathcal{T}^{(n)}(\mathbf{p};\Lambda_n) =
  \Lambda_n\mathcal{T}^{(n)}(0;1) + \mathcal{T}^{(n)}(\mathbf{p};0)\,.
\end{align}
The full tensor flow with transverse-traceless external legs is then
contracted with $\mathcal{T}^{(n)}(0;1)$ or
$\mathcal{T}^{(n)}(\mathbf{p};0)$ in order to yield scalar expressions
that are related to the flow of $\Lambda_n$ or $G_n$, respectively.
In particular, we denote the contraction of the RHS with
$\mathcal{T}^{(n)}(0,1)$ and $\mathcal{T}^{(n)}(\mathbf{p},0)/{\bf p}^2$ by
\begin{small}$\text{Flow}^{(n)}_{\Lambda}$\end{small}
and \begin{small}$\text{Flow}^{(n)}_{G}$\end{small}, respectively. The
factor of $1/{\bf p}^2$ in the definition of the flow for the
gravitational coupling accounts for the fact that the corresponding
tensor projector is proportional to $p^2$.  For convenience, the
definition of \begin{small}$\text{Flow}^{(n)}$\end{small} also
includes the factor \begin{small}$\prod_i
  Z_h^{-1/2}(p_i)$\end{small}. For the graviton three-point function
these objects take the generic form
\begin{align}\label{e:flowint}
  \text{Flow}^{(3)}_{\Lambda/G}= \int_q \left(\dot{r}(q) -
    \eta_{\phi_i}(q^2)r(q) \right)
  F_{{\phi_i},\Lambda/G}(\mathbf{p},q,G_{n},\Lambda_n)\,,
\end{align}
where $n\in\{3,4,5\}$ and a sum over species of fields $\phi_i$ is
understood. Furthermore, $r$ is the shape function of the
regulator. The contributions encoded in $F_{\phi_i}$ originate from
the diagrams displayed in \autoref{fig:flow_three_point}. Note that
\eqref{e:flowint} is only a function of the anomalous dimension
\begin{align}
 \eta_{\phi_i}(p^2) := -\dot Z_{\phi_i}(p^2)/Z_{\phi_i}(p^2) \, ,
\end{align}
since all wave function renormalisations $Z_{\phi_i}$ drop out.  The
expressions for the flow of the three-point function still depend on
the external momenta $\mathbf{p}=(p_1,p_2,p_3)$, where $p_3$ can be
eliminated using momentum conservation. Therefore, the kinematic
degrees of freedom can be parametrised by the absolute values of the
remaining two momenta $|p_1|$, $|p_2|$ and the angle $\vartheta_{12}$
between them. For the proof of locality we work with the most general
momentum configuration. For the
flows \begin{small}$\text{Flow}^{(3)}_{\Lambda/G}$\end{small} the
maximally symmetric momentum configuration is used,
\begin{align}
  p:=|p_1|=|p_2|	&	& 	\vartheta_{12}=2\pi/3\,.
\end{align}
In summary, we have specified a projection procedure for the spacetime
indices and a kinematic configuration for the graviton field
momenta. It remains to
relate \begin{small}$\text{Flow}^{(3)}_{G}$\end{small} and
\begin{small}$\text{Flow}^{(3)}_{\Lambda}$\end{small} to the flow of
the couplings $G_{3}$ and $\Lambda_{3}$, respectively. The flow of
$G_3$ is most conveniently isolated by evaluating the projected flow
at two momentum scales $p=k$ and $p=0$ and subtracting the
results. For the dimensionless coupling $g_3=k^2 G_3$ we obtain
\begin{align} \label{eq:flow-g} \dot g_3 =& (2+3\eta_h(k^2))g_3
  -\frac{24}{19} (\eta_h(k^2)-\eta_h(0))
  \lambda_3 g_3\notag\\
  &+ 2 \, \CN_g \, \sqrt{g_3} \, k
  \left(\text{Flow}^{(3)}_G(k^2)-\text{Flow}^{(3)}_G(0)\right)\,,
\end{align}
with a normalisation factor
$\CN_g^{-1}:=\mathcal{T}^{(3)}(k;0)\circ\Pi_{\text{TT}}^3\circ\mathcal{T}^{(3)}
(k;0)$, where $\circ$ denotes the pairwise contraction of
indices. Another possibility is the evaluation with a $p^2$-derivative
at $p=0$. This procedure is less accurate in approximating the
momentum dependence of the flow. On the other hand, it allows for an
analytic flow equation for the couplings $G_3$. The difference between
these momentum projections is discussed below. The flow of the
dimensionless coupling $\lambda_3=\Lambda_3/k^{2}$ is obtained by
evaluating the flow at $p=0$, which leads to
\begin{align}\label{eq:flow-l}
  \dot \lambda_3 = \left(\frac32\eta_h(0)-1-\frac{\dot g_3}{2
g_3}\right)\lambda_3 + \frac{\CN_\lambda}{\sqrt{g_3}}\text{Flow}^{(3)}_\Lambda(0)\,,
\end{align}
with $\CN_\lambda^{-1} := \mathcal{T}^{(3)} (0;1)\circ\Pi_ {
  \text{TT}}^3\circ\mathcal{T}^{(3)}(0;1)$.  The setup is complemented
by flow equations for the graviton mass parameter $\mu =
-2\Lambda_2/k^2$, the fully momentum dependent anomalous dimensions
$\eta_{\phi_i}(p^2)$ and the coupling of the one-point function
$\lambda_1/\sqrt{g_1}=\Lambda_{1} G_1^{-1/2}/k^{3}$. The
  flows for the couplings $\mu$ and $\lambda_1/\sqrt{g_1}$ are
  extracted from the graviton two- and one-point functions at
  vanishing external momenta, respectively.  The anomalous dimensions
  $\eta_{\phi_i}(p^2)$ are solutions to Fredholm integral equations,
  extracted from the two-point
  functions, see \cite{Christiansen:2014raa}. \\[-1ex]

\begin{figure}[t]
\includegraphics[width=4.25cm]{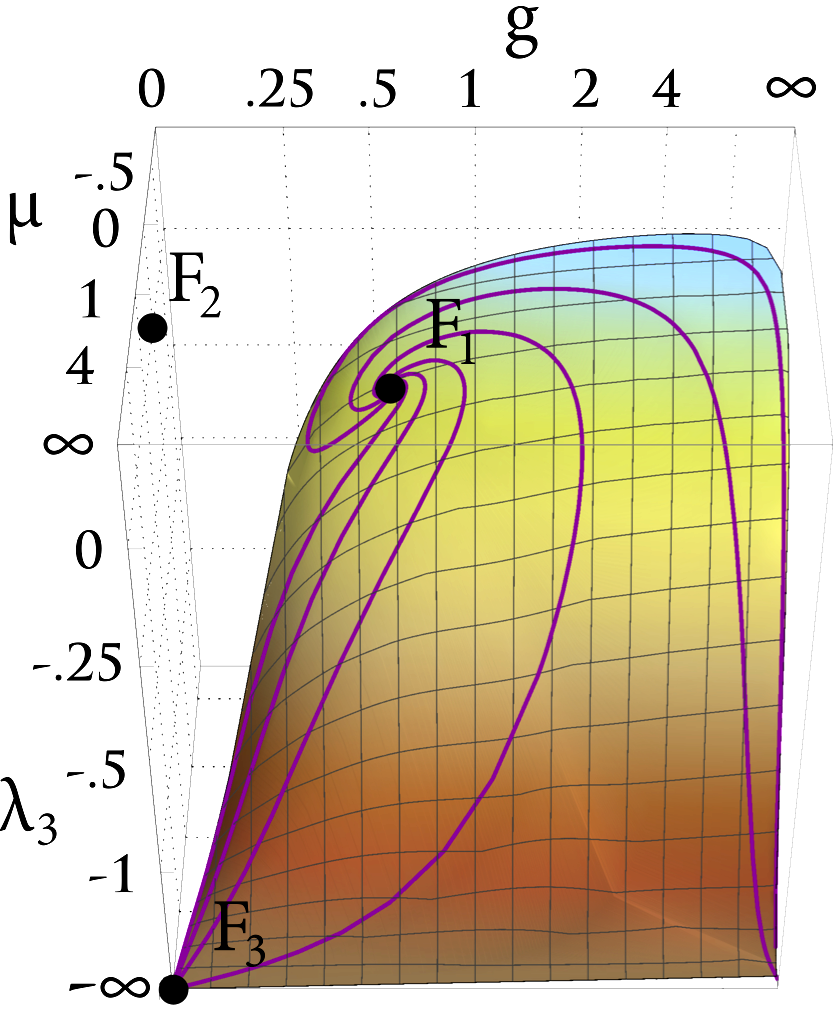}
\includegraphics[width=4.25cm]{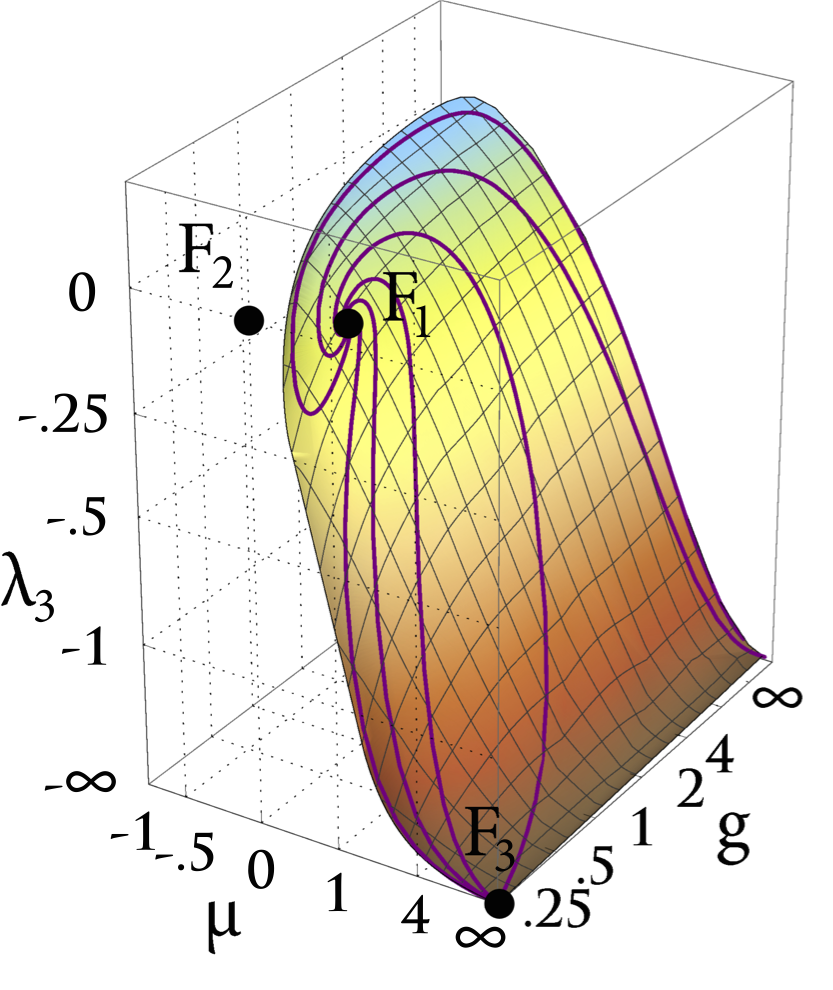}
\caption{%
  Phase diagram for the couplings $g$, $\lambda_3$ and $\mu$ in two
  different views.  The phase diagram was calculated using the
  analytic equations \eqref{eq:flows}. The system exhibits a
  non-trivial UV fixed point, $F_1$, with two attractive and one
  repulsive direction. The Gaussian fixed point and a non-trivial IR
  fixed point are denoted as $F_2$ and $F_3$, respectively. The set of
  trajectories that approach $F_1$ constitutes two-dimensional UV
  critical hypersurface represented in gradient colours.}
\label{fig:phasediag_no_eta}
\end{figure}

{\it Proof of Locality - } In order to prove \eq{eq:Gninfty} for the
three point function, an arbitrary kinematic configuration is used and
parametrised by $|p_1|$, $|p_2|$ and the angle $\vartheta_{12}$.  The
large momentum limit is then characterised by $|p_1|=|p_2|=p \to
\infty$. Simple power counting of the momentum structure of the flow
leads to the na\"{\i}ve expectation that \begin{small}$\lim_{{p}/k \to
    \infty} \text{Flow}^{(3)}_{G} \sim p^2$\end{small}. In this case
the ratio in \eq{eq:Gninfty} would tend to a constant. However, an
analytic asymptotic expansion around $p = \infty$ shows that the
$p^2$-contribution vanishes identically in the large-momentum limit by
non-trivial cancellations between all diagrams in
\autoref{fig:flow_three_point}. As a
consequence, \begin{small}$\lim_{p/k \rightarrow \infty}
  \text{Flow}^{(3)}_{G}$\end{small} tends to a constant and the ratio
in \eqref{eq:Gninfty} vanishes. This is valid for all values of the
angle $\vartheta_{12}$, i.e.\ for all kinematic configurations. For an
explicit example see \autoref{fig:locality} for the symmetric momentum
configuration. \autoref{fig:locality} further displays that
\eqref{eq:Gninfty} is also satisfied by the graviton two-point
function, see also \cite{Christiansen:2012rx,Christiansen:2014raa}. We
conclude that locality is always satisfied by the flows of two- and
three-point functions.  We emphasise again that it is indispensable
that all external momenta are taken to infinity. Indeed, for
configurations with mixed UV-IR limit equation \eqref{eq:Gninfty} does
not hold.
\\[-1ex]

{\it UV Fixed Point - }
Fixed points are defined by vanishing flows of all dimensionless
dynamical couplings, that is $g_3$, $\lambda_3$ and $\mu$ in the
present setup. Most importantly, we find a UV fixed point with
one irrelevant direction that is approximately directed along the
$\lambda_3$-axis.

The following results are obtained with the regulator 
$R_\phi(x) = \Gamma^{(\phi\phi)}_k|_{\mu=0}(x)\, r(x)$
where $x r(x) =(1-x)\theta(1-x)$. Moreover, we identify
$\lambda_3\equiv\lambda_4\equiv\lambda_5$ in order to close the flow
equations, and use the notation $g:= g_3$. The UV fixed point
described below is obtained with the finite difference procedure,
leading to the flow equations \eqref{eq:flow-g} and \eqref{eq:flow-l},
as well as the one for $\mu$ already presented in
\cite{Christiansen:2014raa}. The anomalous dimensions are evaluated with 
their full momentum dependence. The fixed
point values read
\begin{align}\label{eq:uvfp}
 (g^*,\mu^*,\lambda_3^*)= (0.66,-0.59,0.11)\,,
\end{align}
with the critical exponents $\theta_1$, $\theta_2$ and $\theta_3$ given by
\begin{align}\label{eq:critexp}
 (\theta_{1/2},\theta_3)=(-1.4\pm4.1\, i ,14)\,.
\end{align}
As already mentioned above, the UV fixed point \eqref{eq:uvfp} has the
interesting property that it is not fully UV attractive: it exhibits
two relevant and one irrelevant direction. In \eqref{eq:critexp}, this
is reflected by two critical exponents with negative real parts,
$\theta_1$ and $\theta_2$, and one with positive real part,
$\theta_3$. The irrelevant direction of the UV fixed point
\eqref{eq:uvfp} is approximately directed along the $\lambda_3$ axis. The
critical exponents corresponding to the UV relevant directions of the
fixed points are complex, which accounts for a spiral behaviour of
RG-trajectories in the vicinity of the UV fixed point. Note, that
$\theta_3$ in \eqref{eq:critexp} is one order of magnitude larger than
$\theta_{1}$ and $\theta_2$. This kind of instability of critical
exponents was also found in \cite{Falls:2014tra} within
$f(R)$-gravity. There, a convergence of the critical exponents to
smaller values was observed after the inclusion of higher order
operators, i.e. higher powers $R^n$. Similarly, we expect $\theta_3$
to become smaller, if dynamical couplings $g_4$ and $\lambda_4$ are
included.
\begin{table}[t]
  \caption{Properties of the UV fixed point for different momentum parametrisations, 
    namely using a finite difference of the flow and a derivative at $p=0$. The values 
    acquired with the latter correspond to the analytic equations given in (\ref{eq:flows}). 
    Note, that $\lambda_1/\sqrt{g_1}$ is a non-dynamical background coupling 
    originating from the graviton one-point function.}
 \label{t:results}
 \vspace{5pt}
 \centering
 \includegraphics[width=8.65cm]{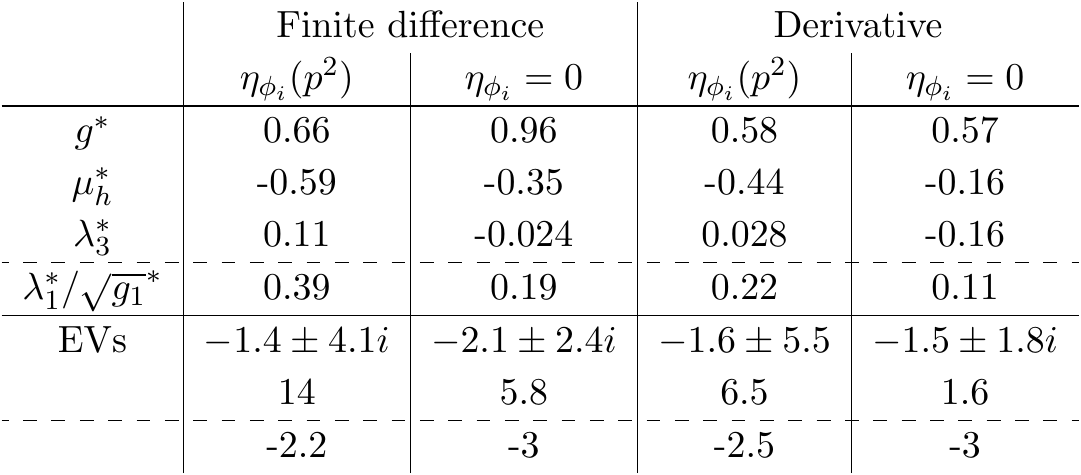}
\end{table}
Finally, we have checked numerically that the momentum dependencies of
our ansatz for the vertex dressing and that 
of \begin{small}$\text{Flow}^{(2)}$\end{small} and
\begin{small}$\text{Flow}^{(3)}_G$\end{small} are in very good
agreement. Therefore, a momentum independent $G_3$ is a valid
approximation over the whole momentum range. In particular, the full
momentum dependence of the anomalous dimensions was found to model
very accurately the higher order $p$-dependence of
\begin{small}$\text{Flow}^{(3)}_G$\end{small}.
\\[-1ex]

{\it Global Phase Diagram and Analytic Flow Equations -} 
The flow equation \eqref{eq:flow-g} does not have a closed analytic form. 
However, for a more accessible presentation, analytic flow equations are favourable.
An analytic expression for $\dot g$ is obtained by taking a derivative of 
\begin{small}$\text{Flow}^{(3)}_G$\end{small} with respect to $p^2$ at
$p=0$. We stress, that this method is considerably less accurate in
modelling the momentum dependence of the flow. Nonetheless, the
resulting analytic flow equation for $g$ shares the main features with
\eqref{eq:flow-g}. This even holds for all anomalous dimensions set
to zero, $\eta_{\phi_i}=0$. \autoref{t:results} displays the
properties of the non-trivial fixed point as obtained from the
different methods. 

The analytic flow equations for the presented
vertex flow of gravity with $\eta_{\phi_i}=0$ are given by 
\begin{small}
\begin{align}\nonumber 
  \dot g =&2g+\frac{8 g^2}{19\pi}\Bigg(\frac{584 \lambda_3^3- 910
    \lambda_3^2+445 \lambda_3-\frac{299}{4}}{15 (\mu +1)^5}
  -\frac{47}{8(\mu +1)^2}	\\
  \nonumber  
  -\frac58&+\frac{864 \lambda_3^3+133 \lambda_3^2-112
    \lambda_3+\frac{49}4}{6 (\mu +1)^4} -\frac{60 \lambda_3^2 -58
    \lambda_3-15}{6 (\mu +1)^3}\Bigg),\\
    \nonumber 
\dot \lambda_3=& -\left(1+\frac{\dot g}{2g}\right)
\lambda_3+\frac{g}{\pi} \Bigg(\frac{2 \lambda_3^3-4 \lambda_3^2+
3 \lambda_3-\frac{11}{20}}{(\mu +1)^4}\\\nonumber 
&\hspace{2.1cm} -4\frac{4 \lambda_3^2 - \lambda_3}{(\mu +1)^3}+\frac{1-
3 \lambda_3}{(\mu +1)^2}+\frac{6}{5}\Bigg)\,,\\
\nonumber 
\dot \mu=& -2\mu+\frac{2g}{\pi}\left(\frac{16 \lambda_3^2-
8\lambda_3+\frac7{4}}{3 (\mu +1)^3}+\frac{2 \lambda_3-1}{ (\mu 
+1)^2}-1\right)\,,\\
\partial_t &\left(\frac{\lambda_1}{\sqrt{g_1}}\right) = -3\frac{
  \lambda_1}{\sqrt{g_1}}+\frac{\sqrt{g}}{2\pi}\left(\frac{1}{(\mu
    +1)^{2}}+\frac43 \right)\,. \label{eq:flows}
\end{align}
\end{small}
\autoref{fig:phasediag_no_eta} shows the phase diagram for the
couplings $(g,\mu,\lambda_3)$ as calculated from
\eqref{eq:flows}.  The purple lines are trajectories along the flow
that terminate at the non-trivial UV fixed point $F_1$. The set of all
trajectories constitutes the two-dimensional critical hypersurface
represented in gradient colours. In the IR, the trajectories flow
towards $F_3=(0,\infty,-\infty)$ or, alternatively, towards
$(\infty,\infty,-\infty)$.  The IR fixed point $F_3$ was also observed
in \cite{Christiansen:2014raa}. In the vicinity of $F_3$ all couplings
scale classically since for $\mu \to \infty$ the loop contributions to
the flow tend to zero. Neither the trivial Gaussian fixed point $F_2$,
nor the third, non-trivial IR fixed point, which was first found in
\cite{Christiansen:2014raa} and is located at $(0,-1,\infty)$, are
reached by any UV-finite trajectory in the presented setup. For the
latter, this is expected to change if the vertices are expanded about
a non-flat background \cite{Falls:2014zba}. A detailed IR analysis is
beyond the scope of this letter and is postponed to future work. \\[-1ex]

{\it Summary - } We have presented the first genuine
  calculation of dynamical gravitational couplings based on a vertex
  flow. The dynamical parameters are the graviton-mass
  parameter $\mu$, the Newton coupling $g$ derived from the graviton
  three-point function, and the coupling $\lambda_3$ of its
  momentum-independent part. The full momentum-dependence of the
propagators is encoded in $\eta_h(p^2)$ and 
$\eta_c(p^2)$. The flows of these quantities constitute a minimally
self-consistent truncation of the system of dynamical couplings in
quantum gravity. In the UV we find a fixed point with two relevant and
one irrelevant direction, the latter being approximately directed
along the $\lambda_3$-axis. This hints at a finite dimensional
critical UV-surface, and supports the asymptotic safety scenario.

We have introduced the property of momentum locality for vertex flows.
It is suggestive that this property is a necessary requirement for
local quantum field theories. We have shown that it is non-trivially
realised for the graviton two- and three-point functions, being linked
to diffeomorphism invariance.

Currently, we extend the present work to improved approximations
including higher order correlation functions and non-trivial backgrounds.
This allows to evaluate the convergence of the present
expansion scheme and its dependence on the expansion point.
Non-trivial backgrounds are expected to play an important
r$\hat{\rm o}$le for the fate of the IR fixed point at $\mu=-1$.
Higher order vertices give further insights into
the important feature of 
momentum locality.\\[-1.5ex]

\noindent {\bf Acknowledgements} We thank A.~Eichhorn, K.~Falls, H.~Gies,
A.~Rodigast and C.~Wetterich for discussions. NC acknowledges funding
from the HGSFP, MR from IMPRS-PTFS, and BK from GRK 1523/2, and Wi
777/11-1. This work is supported by the Helmholtz Alliance HA216/EMMI
and by ERC-AdG-290623.

 \bibliography{flatgravity}

\begin{thebibliography}{18}%
\makeatletter
\providecommand \@ifxundefined [1]{%
 \@ifx{#1\undefined}
}%
\providecommand \@ifnum [1]{%
 \ifnum #1\expandafter \@firstoftwo
 \else \expandafter \@secondoftwo
 \fi
}%
\providecommand \@ifx [1]{%
 \ifx #1\expandafter \@firstoftwo
 \else \expandafter \@secondoftwo
 \fi
}%
\providecommand \natexlab [1]{#1}%
\providecommand \enquote  [1]{``#1''}%
\providecommand \bibnamefont  [1]{#1}%
\providecommand \bibfnamefont [1]{#1}%
\providecommand \citenamefont [1]{#1}%
\providecommand \href@noop [0]{\@secondoftwo}%
\providecommand \href [0]{\begingroup \@sanitize@url \@href}%
\providecommand \@href[1]{\@@startlink{#1}\@@href}%
\providecommand \@@href[1]{\endgroup#1\@@endlink}%
\providecommand \@sanitize@url [0]{\catcode `\\12\catcode `\$12\catcode
  `\&12\catcode `\#12\catcode `\^12\catcode `\_12\catcode `\%12\relax}%
\providecommand \@@startlink[1]{}%
\providecommand \@@endlink[0]{}%
\providecommand \url  [0]{\begingroup\@sanitize@url \@url }%
\providecommand \@url [1]{\endgroup\@href {#1}{\urlprefix }}%
\providecommand \urlprefix  [0]{URL }%
\providecommand \Eprint [0]{\href }%
\providecommand \doibase [0]{http://dx.doi.org/}%
\providecommand \selectlanguage [0]{\@gobble}%
\providecommand \bibinfo  [0]{\@secondoftwo}%
\providecommand \bibfield  [0]{\@secondoftwo}%
\providecommand \translation [1]{[#1]}%
\providecommand \BibitemOpen [0]{}%
\providecommand \bibitemStop [0]{}%
\providecommand \bibitemNoStop [0]{.\EOS\space}%
\providecommand \EOS [0]{\spacefactor3000\relax}%
\providecommand \BibitemShut  [1]{\csname bibitem#1\endcsname}%
\let\auto@bib@innerbib\@empty
\bibitem [{\citenamefont {Weinberg}(1979)}]{Weinberg:1980gg}%
  \BibitemOpen
  \bibfield  {author} {\bibinfo {author} {\bibfnamefont {S.}~\bibnamefont
  {Weinberg}},\ }\href@noop {} {\bibfield  {journal} {\bibinfo  {journal}
  {General Relativity: An Einstein centenary survey, Eds. Hawking, S.W.,
  Israel, W; Cambridge University Press}\ ,\ \bibinfo {pages} {790}} (\bibinfo
  {year} {1979})}\BibitemShut {NoStop}%
\bibitem [{\citenamefont {Reuter}(1998)}]{Reuter:1996cp}%
  \BibitemOpen
  \bibfield  {author} {\bibinfo {author} {\bibfnamefont {M.}~\bibnamefont
  {Reuter}},\ }\href {\doibase 10.1103/PhysRevD.57.971} {\bibfield  {journal}
  {\bibinfo  {journal} {Phys. Rev.}\ }\textbf {\bibinfo {volume} {D57}},\
  \bibinfo {pages} {971} (\bibinfo {year} {1998})},\ \Eprint
  {http://arxiv.org/abs/hep-th/9605030} {arXiv:hep-th/9605030} \BibitemShut
  {NoStop}%
\bibitem [{\citenamefont {Christiansen}\ \emph
  {et~al.}(2014{\natexlab{a}})\citenamefont {Christiansen}, \citenamefont
  {Litim}, \citenamefont {Pawlowski},\ and\ \citenamefont
  {Rodigast}}]{Christiansen:2012rx}%
  \BibitemOpen
  \bibfield  {author} {\bibinfo {author} {\bibfnamefont {N.}~\bibnamefont
  {Christiansen}}, \bibinfo {author} {\bibfnamefont {D.~F.}\ \bibnamefont
  {Litim}}, \bibinfo {author} {\bibfnamefont {J.~M.}\ \bibnamefont
  {Pawlowski}}, \ and\ \bibinfo {author} {\bibfnamefont {A.}~\bibnamefont
  {Rodigast}},\ }\href {\doibase 10.1016/j.physletb.2013.11.025} {\bibfield
  {journal} {\bibinfo  {journal} {Phys.Lett.}\ }\textbf {\bibinfo {volume}
  {B728}},\ \bibinfo {pages} {114} (\bibinfo {year} {2014}{\natexlab{a}})},\
  \Eprint {http://arxiv.org/abs/1209.4038} {arXiv:1209.4038 [hep-th]}
  \BibitemShut {NoStop}%
\bibitem [{\citenamefont {Christiansen}\ \emph
  {et~al.}(2014{\natexlab{b}})\citenamefont {Christiansen}, \citenamefont
  {Knorr}, \citenamefont {Pawlowski},\ and\ \citenamefont
  {Rodigast}}]{Christiansen:2014raa}%
  \BibitemOpen
  \bibfield  {author} {\bibinfo {author} {\bibfnamefont {N.}~\bibnamefont
  {Christiansen}}, \bibinfo {author} {\bibfnamefont {B.}~\bibnamefont {Knorr}},
  \bibinfo {author} {\bibfnamefont {J.~M.}\ \bibnamefont {Pawlowski}}, \ and\
  \bibinfo {author} {\bibfnamefont {A.}~\bibnamefont {Rodigast}},\ }\href@noop
  {} {\  (\bibinfo {year} {2014}{\natexlab{b}})},\ \Eprint
  {http://arxiv.org/abs/1403.1232} {arXiv:1403.1232 [hep-th]} \BibitemShut
  {NoStop}%
\bibitem [{\citenamefont {Donkin}\ and\ \citenamefont
  {Pawlowski}(2012)}]{Donkin:2012ud}%
  \BibitemOpen
  \bibfield  {author} {\bibinfo {author} {\bibfnamefont {I.}~\bibnamefont
  {Donkin}}\ and\ \bibinfo {author} {\bibfnamefont {J.~M.}\ \bibnamefont
  {Pawlowski}},\ }\href@noop {} {\  (\bibinfo {year} {2012})},\ \Eprint
  {http://arxiv.org/abs/1203.4207} {arXiv:1203.4207 [hep-th]} \BibitemShut
  {NoStop}%
\bibitem [{\citenamefont {Falls}\ \emph {et~al.}(2014)\citenamefont {Falls},
  \citenamefont {Litim}, \citenamefont {Nikolakopoulos},\ and\ \citenamefont
  {Rahmede}}]{Falls:2014tra}%
  \BibitemOpen
  \bibfield  {author} {\bibinfo {author} {\bibfnamefont {K.}~\bibnamefont
  {Falls}}, \bibinfo {author} {\bibfnamefont {D.~F.}\ \bibnamefont {Litim}},
  \bibinfo {author} {\bibfnamefont {K.}~\bibnamefont {Nikolakopoulos}}, \ and\
  \bibinfo {author} {\bibfnamefont {C.}~\bibnamefont {Rahmede}},\ }\href@noop
  {} {\  (\bibinfo {year} {2014})},\ \Eprint {http://arxiv.org/abs/1410.4815}
  {arXiv:1410.4815 [hep-th]} \BibitemShut {NoStop}%
\bibitem [{\citenamefont {Codello}\ \emph {et~al.}(2008)\citenamefont
  {Codello}, \citenamefont {Percacci},\ and\ \citenamefont
  {Rahmede}}]{Codello:2007bd}%
  \BibitemOpen
  \bibfield  {author} {\bibinfo {author} {\bibfnamefont {A.}~\bibnamefont
  {Codello}}, \bibinfo {author} {\bibfnamefont {R.}~\bibnamefont {Percacci}}, \
  and\ \bibinfo {author} {\bibfnamefont {C.}~\bibnamefont {Rahmede}},\ }\href
  {\doibase 10.1142/S0217751X08038135} {\bibfield  {journal} {\bibinfo
  {journal} {Int. J. Mod. Phys.}\ }\textbf {\bibinfo {volume} {A23}},\ \bibinfo
  {pages} {143} (\bibinfo {year} {2008})},\ \Eprint
  {http://arxiv.org/abs/0705.1769} {arXiv:0705.1769 [hep-th]} \BibitemShut
  {NoStop}%
\bibitem [{\citenamefont {Machado}\ and\ \citenamefont
  {Saueressig}(2008)}]{Machado:2007ea}%
  \BibitemOpen
  \bibfield  {author} {\bibinfo {author} {\bibfnamefont {P.~F.}\ \bibnamefont
  {Machado}}\ and\ \bibinfo {author} {\bibfnamefont {F.}~\bibnamefont
  {Saueressig}},\ }\href {\doibase 10.1103/PhysRevD.77.124045} {\bibfield
  {journal} {\bibinfo  {journal} {Phys. Rev.}\ }\textbf {\bibinfo {volume}
  {D77}},\ \bibinfo {pages} {124045} (\bibinfo {year} {2008})},\ \Eprint
  {http://arxiv.org/abs/0712.0445} {arXiv:0712.0445 [hep-th]} \BibitemShut
  {NoStop}%
\bibitem [{\citenamefont {Benedetti}\ \emph {et~al.}(2009)\citenamefont
  {Benedetti}, \citenamefont {Machado},\ and\ \citenamefont
  {Saueressig}}]{Benedetti:2009rx}%
  \BibitemOpen
  \bibfield  {author} {\bibinfo {author} {\bibfnamefont {D.}~\bibnamefont
  {Benedetti}}, \bibinfo {author} {\bibfnamefont {P.~F.}\ \bibnamefont
  {Machado}}, \ and\ \bibinfo {author} {\bibfnamefont {F.}~\bibnamefont
  {Saueressig}},\ }\href {\doibase 10.1142/S0217732309031521} {\bibfield
  {journal} {\bibinfo  {journal} {Mod. Phys. Lett.}\ }\textbf {\bibinfo
  {volume} {A24}},\ \bibinfo {pages} {2233} (\bibinfo {year} {2009})},\ \Eprint
  {http://arxiv.org/abs/0901.2984} {arXiv:0901.2984 [hep-th]} \BibitemShut
  {NoStop}%
\bibitem [{\citenamefont {Manrique}\ \emph {et~al.}(2011)\citenamefont
  {Manrique}, \citenamefont {Reuter},\ and\ \citenamefont
  {Saueressig}}]{Manrique:2010am}%
  \BibitemOpen
  \bibfield  {author} {\bibinfo {author} {\bibfnamefont {E.}~\bibnamefont
  {Manrique}}, \bibinfo {author} {\bibfnamefont {M.}~\bibnamefont {Reuter}}, \
  and\ \bibinfo {author} {\bibfnamefont {F.}~\bibnamefont {Saueressig}},\
  }\href {\doibase 10.1016/j.aop.2010.11.006} {\bibfield  {journal} {\bibinfo
  {journal} {Annals Phys.}\ }\textbf {\bibinfo {volume} {326}},\ \bibinfo
  {pages} {463} (\bibinfo {year} {2011})},\ \Eprint
  {http://arxiv.org/abs/1006.0099} {arXiv:1006.0099 [hep-th]} \BibitemShut
  {NoStop}%
\bibitem [{\citenamefont {Becker}\ and\ \citenamefont
  {Reuter}(2014)}]{Becker:2014qya}%
  \BibitemOpen
  \bibfield  {author} {\bibinfo {author} {\bibfnamefont {D.}~\bibnamefont
  {Becker}}\ and\ \bibinfo {author} {\bibfnamefont {M.}~\bibnamefont
  {Reuter}},\ }\href {\doibase 10.1016/j.aop.2014.07.023} {\bibfield  {journal}
  {\bibinfo  {journal} {Annals Phys.}\ }\textbf {\bibinfo {volume} {350}},\
  \bibinfo {pages} {225} (\bibinfo {year} {2014})},\ \Eprint
  {http://arxiv.org/abs/1404.4537} {arXiv:1404.4537 [hep-th]} \BibitemShut
  {NoStop}%
\bibitem [{\citenamefont {Nagy}\ \emph {et~al.}(2012)\citenamefont {Nagy},
  \citenamefont {Krizsan},\ and\ \citenamefont {Sailer}}]{Nagy:2012rn}%
  \BibitemOpen
  \bibfield  {author} {\bibinfo {author} {\bibfnamefont {S.}~\bibnamefont
  {Nagy}}, \bibinfo {author} {\bibfnamefont {J.}~\bibnamefont {Krizsan}}, \
  and\ \bibinfo {author} {\bibfnamefont {K.}~\bibnamefont {Sailer}},\ }\href
  {\doibase 10.1007/JHEP07(2012)102} {\bibfield  {journal} {\bibinfo  {journal}
  {JHEP}\ }\textbf {\bibinfo {volume} {1207}},\ \bibinfo {pages} {102}
  (\bibinfo {year} {2012})},\ \Eprint {http://arxiv.org/abs/1203.6564}
  {arXiv:1203.6564 [hep-th]} \BibitemShut {NoStop}%
\bibitem [{\citenamefont {Benedetti}(2013)}]{Benedetti:2013jk}%
  \BibitemOpen
  \bibfield  {author} {\bibinfo {author} {\bibfnamefont {D.}~\bibnamefont
  {Benedetti}},\ }\href {\doibase 10.1209/0295-5075/102/20007} {\bibfield
  {journal} {\bibinfo  {journal} {Europhys.Lett.}\ }\textbf {\bibinfo {volume}
  {102}},\ \bibinfo {pages} {20007} (\bibinfo {year} {2013})},\ \Eprint
  {http://arxiv.org/abs/1301.4422} {arXiv:1301.4422 [hep-th]} \BibitemShut
  {NoStop}%
\bibitem [{\citenamefont {Pawlowski}(2003)}]{Pawlowski:2003sk}%
  \BibitemOpen
  \bibfield  {author} {\bibinfo {author} {\bibfnamefont {J.~M.}\ \bibnamefont
  {Pawlowski}},\ }\href@noop {} {\  (\bibinfo {year} {2003})},\ \Eprint
  {http://arxiv.org/abs/hep-th/0310018} {arXiv:hep-th/0310018 [hep-th]}
  \BibitemShut {NoStop}%
\bibitem [{\citenamefont {Pawlowski}(2007)}]{Pawlowski:2005xe}%
  \BibitemOpen
  \bibfield  {author} {\bibinfo {author} {\bibfnamefont {J.~M.}\ \bibnamefont
  {Pawlowski}},\ }\href {\doibase 10.1016/j.aop.2007.01.007} {\bibfield
  {journal} {\bibinfo  {journal} {Annals Phys.}\ }\textbf {\bibinfo {volume}
  {322}},\ \bibinfo {pages} {2831} (\bibinfo {year} {2007})},\ \Eprint
  {http://arxiv.org/abs/hep-th/0512261} {arXiv:hep-th/0512261 [hep-th]}
  \BibitemShut {NoStop}%
\bibitem [{\citenamefont {Folkerts}\ \emph {et~al.}(2012)\citenamefont
  {Folkerts}, \citenamefont {Litim},\ and\ \citenamefont
  {Pawlowski}}]{Folkerts:2011jz}%
  \BibitemOpen
  \bibfield  {author} {\bibinfo {author} {\bibfnamefont {S.}~\bibnamefont
  {Folkerts}}, \bibinfo {author} {\bibfnamefont {D.~F.}\ \bibnamefont {Litim}},
  \ and\ \bibinfo {author} {\bibfnamefont {J.~M.}\ \bibnamefont {Pawlowski}},\
  }\href {\doibase 10.1016/j.physletb.2012.02.002} {\bibfield  {journal}
  {\bibinfo  {journal} {Phys.Lett.}\ }\textbf {\bibinfo {volume} {B709}},\
  \bibinfo {pages} {234} (\bibinfo {year} {2012})},\ \Eprint
  {http://arxiv.org/abs/1101.5552} {arXiv:1101.5552 [hep-th]} \BibitemShut
  {NoStop}%
\bibitem [{\citenamefont {Fischer}\ and\ \citenamefont
  {Pawlowski}(2009)}]{Fischer:2009tn}%
  \BibitemOpen
  \bibfield  {author} {\bibinfo {author} {\bibfnamefont {C.~S.}\ \bibnamefont
  {Fischer}}\ and\ \bibinfo {author} {\bibfnamefont {J.~M.}\ \bibnamefont
  {Pawlowski}},\ }\href {\doibase 10.1103/PhysRevD.80.025023} {\bibfield
  {journal} {\bibinfo  {journal} {Phys. Rev.}\ }\textbf {\bibinfo {volume}
  {D80}},\ \bibinfo {pages} {025023} (\bibinfo {year} {2009})},\ \Eprint
  {http://arxiv.org/abs/0903.2193} {arXiv:0903.2193 [hep-th]} \BibitemShut
  {NoStop}%
\bibitem [{\citenamefont {Falls}(2014)}]{Falls:2014zba}%
  \BibitemOpen
  \bibfield  {author} {\bibinfo {author} {\bibfnamefont {K.}~\bibnamefont
  {Falls}},\ }\href@noop {} {\  (\bibinfo {year} {2014})},\ \Eprint
  {http://arxiv.org/abs/1408.0276} {arXiv:1408.0276 [hep-th]} \BibitemShut
  {NoStop}%
\end{thebibliography}%

\end{document}